\documentclass[preprint,amssymb,amsmath,floatfix,twocolumn,3p]{elsarticle}
\newcommand{\be}{\begin{equation}}
\newcommand{\ee}{\end{equation}}
\newcommand{\bea}{\begin{eqnarray}}
\newcommand{\eea}{\end{eqnarray}}

\usepackage{graphicx}
\usepackage{epsfig}
\usepackage{mathtools}
\usepackage{color}

\begin{document}

\title{{Many-body} braiding phases in a rotating strongly correlated photon gas}

\author[BEC]{R. O. Umucal\i lar\corref{cor1}\fnref{fn1}}
\ead{rifatonur.umucalilar@ua.ac.be}
\author[BEC]{I. Carusotto}
\address[BEC]{INO-CNR BEC Center and Dipartimento di Fisica, Universit\`a di Trento, I-38123 Povo, Italy}
\cortext[cor1]{Corresponding author}
\fntext[fn1]{Present Address: TQC, Universiteit Antwerpen, Universiteitsplein 1, B-2610 Antwerpen, Belgium}

\begin{abstract}
We present a theoretical study of fractional quantum Hall physics in a rotating gas of strongly interacting photons in a single cavity with a large optical nonlinearity. Photons are injected into the cavity by a Laguerre-Gauss laser beam with a non-zero orbital angular momentum. The Laughlin-like few-photon eigenstates appear as sharp resonances in the transmission spectra. Using additional localized repulsive potentials, quasi-holes can be created in the photon gas and then braided around in space: an unambiguous signature of the many-body Berry phase under exchange of two quasi-holes is observed as a spectral shift of the corresponding transmission resonance.
\end{abstract}


\maketitle

\section{Introduction}
Quasi-particles with fractional statistics in a two-dimensional electronic gas under a strong magnetic field in the fractional quantum Hall (FQH) regime are among the most fascinating discoveries of contemporary condensed-matter physics~\cite{Yoshioka} and are raising an ever-growing excitement in view of topological quantum computation applications~\cite{DasSarma}. While the fractional value of their charge has been clearly observed in shot-noise experiments~\cite{dePicciotto}, evidence of the many-body braiding phase under exchange of two quasi-particles \cite{Wu} is still quite elusive~\cite{Camino,Stern}.

In the last decade, nonlinear optical systems have been emerging as an outstanding new platform to study quantum many-body physics in gases of many interacting photons~\cite{ICCC_RMP}: superfluid hydrodynamic effects have been experimentally investigated with unprecedented detail in polariton gases in semiconductor microcavities~\cite{superfl} and an intense experimental effort is being devoted to the generation of strongly correlated states like Mott insulator~\cite{Mott} or Tonks-Girardeau gases~\cite{TGfiber,IC_TG}.
Even though optical vortices in nonlinear optical media have received a great attention since the earliest works in fluids of light~\cite{ICCC_RMP} and the experimental generation of synthetic gauge fields for photons has been recently reported~\cite{synthetic_exp,hafezi_exp,segev_exp}, so far very few works have explored the interplay of the orbital angular momentum of light with strong photon-photon interactions at the single quantum level~\cite{Bose,FQH_photon,Greentree,Hafezi_FQH}.

In this Letter, we theoretically discuss an all-optical set-up where the FQH physics can be explored in a gas of photons. A Laguerre-Gauss laser beam with a non-zero orbital angular momentum is used to inject rotating photons into a single cavity bounded by spherical mirrors, whose curvature provides a harmonic trapping along the plane orthogonal to the cavity axis. In exactly the same way as predicted for ultra-cold atomic clouds~\cite{Fetter,Cooper_review,Dalib_rev}, the close analogy between the Coriolis force in a rotating reference frame and the Lorentz force under a magnetic field anticipates the appearance of strongly correlated quantum Hall liquids for fast enough rotations. In the present photonic case, the required repulsive interactions between photons are provided by a strong $\chi^{(3)}$ optical nonlinearity in the cavity medium.

The main result of this work concerns the appearance of sharp peaks in the transmission spectrum of the cavity, whose origin can be traced back~\cite{IC_TG,FQH_photon} to few photon states with excellent overlap with Laughlin states of FQH physics~\cite{Laughlin}. As compared to previous studies of quantum Hall physics in photon gases~\cite{Bose,FQH_photon,Greentree,Hafezi_FQH}, our proposal does not require sophisticated fabrication techniques to generate the synthetic gauge field for photons~\cite{synthetic_exp,hafezi_exp,segev_exp,synthetic_th}. Even more remarkably, our proposed set-up gives direct access to the
many-body Berry phase~\cite{Berry}, a quantity that is at the core of the anyonic statistics predicted to emerge in FQH systems~\cite{Stern,Halperin,Wilczek_anyon}. A related proposal to measure the many-body Berry phase in rotating ultra-cold atomic clouds appeared in~\cite{Zoller_anyon}. In contrast to interferometrical experiments on electron gases~\cite{Camino,Stern}, neither this proposal nor ours is expected to be subject to fundamental interpretation difficulties arising from competing effects.

\section{Model system}
\begin{figure}[h]
\includegraphics[width = 0.48\columnwidth]{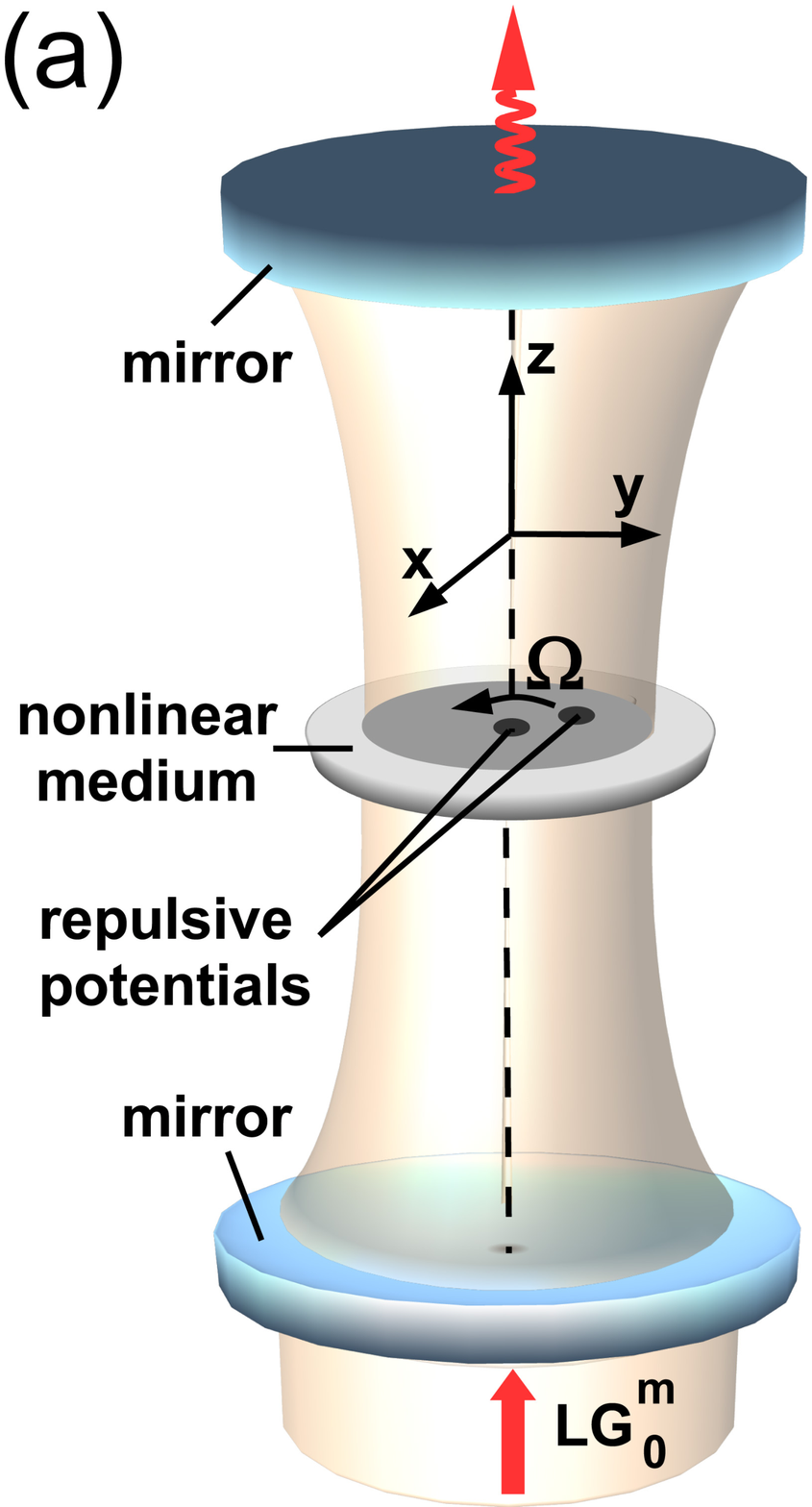}
\includegraphics[width = 0.50\columnwidth]{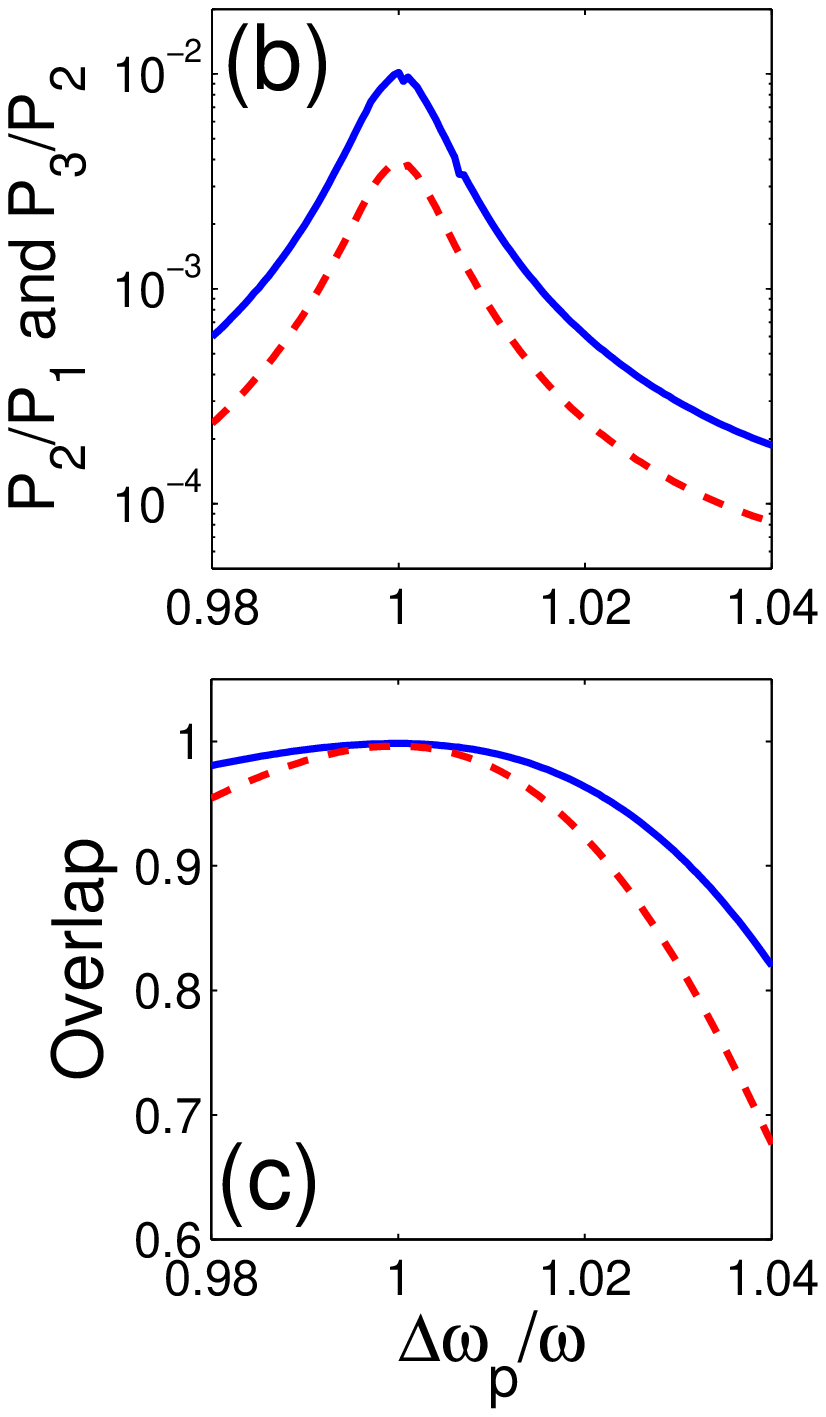}
\caption{(a) Sketch of the experimental setup. (b) Steady-state relative probability of two-particle excitation $P_2/P_1$ (solid line) and of three-particle excitation $P_3/P_2$ (dashed line)  in the absence of repulsive potentials under a monochromatic Laguerre-Gauss pump ${\rm LG}_0^1$ and ${\rm LG}_0^2$, respectively.
(c) Overlap of the two- (solid line) and three-photon (dashed line) amplitudes with the corresponding Laughlin wave function. System parameters: $\Omega/\omega = 1$, $g_{nl}/\ell^2\omega = 4$, $\gamma/\omega = 0.01$, $\ell F/\gamma = 0.1$. The three-body calculation is performed within the LLL approximation.
\label{Fig1}}
\end{figure}

The physical system we are considering is a single optical cavity with cylindrical symmetry consisting of a pair of spherical mirrors and containing a slab of nonlinear medium as sketched in Fig. \ref{Fig1}(a). Transverse modes with a given longitudinal mode number $\mathcal{N}_z$ along the cavity axis $\hat{\mathbf{z}}$ can be described as the eigenstates of an isotropic two-dimensional harmonic oscillator of frequency $\omega=\sqrt{2c^2/R L}$, $L$ being the central distance between the two mirrors and $R$ their radius of curvature.
As the polarization and orbital degrees of freedom are very weakly coupled in actual {cavities} of this kind~\cite{Klaers}, we restrict our model to a single polarization state selected by the polarization of the incident light. The many-body dynamics of cavity photons in the given longitudinal mode can then be described in second quantization terms via the field Hamiltonian~\cite{ICCC_RMP}
\begin{multline}
\mathcal{H}=\int\!d^2\mathbf{r}\,\left\{\frac{\hbar^2}{2m_{ph}}\,\nabla \hat{\Psi}^\dagger(\mathbf{r})\,\nabla \hat{\Psi}(\mathbf{r})+ \right. \\ +\left[\hbar\omega_c+\frac{m_{ph}\,\omega^2\,r^2}{2}+V_{qh}(\mathbf{r},t)\right]\,\hat{\Psi}^\dagger(\mathbf{r})\,\hat{\Psi}(\mathbf{r})+ \\
+\left. \frac{\hbar g_{nl}}{2}\,\hat{\Psi}^\dagger(\mathbf{r})\hat{\Psi}^\dagger(\mathbf{r})
\hat{\Psi}(\mathbf{r})
\hat{\Psi}(\mathbf{r})+ \right. \\
\left. +\hbar F(\mathbf{r},t)\,\hat{\Psi}^\dagger(\mathbf{r}) + \hbar F^*(\mathbf{r},t)\,\hat{\Psi}(\mathbf{r})\right\},
\label{eq:H}
\end{multline}
where the two-dimensional quantum photon field $\hat{\Psi}(\mathbf{r})$ satisfies two-dimensional bosonic commutation rules $[\hat{\Psi}(\mathbf{r}),\hat{\Psi}^\dagger(\mathbf{r}')]=\delta^{(2)}(\mathbf{r}-\mathbf{r}')$. The confinement between the two mirrors is responsible for the finite photon rest frequency $\omega_c=c \pi \mathcal{N}_z /L$ and its mass $m_{ph}=\hbar \omega_c/c^2$ and the mirror curvature provides the harmonic trapping potential of frequency $\omega$~\cite{Klaers}.
Of course, this same Hamiltonian can be used to describe a variety of other configurations, e.g. solid-state planar microcavities with a suitable lateral patterning~\cite{Deppe,Deveaud}, or even hybrid set-ups with a spherical fiber-tip mirror facing a planar DBR mirror~\cite{Thomas}.

The additional potential $V_{qh}(\mathbf{r},t)$ will be taken as a sum of $N_{qh}$ repulsive delta-shaped potentials of strength $V_\circ$ centered at time-dependent positions $\mathbf{r}^{({\rm lab})}_i(t)$, $V_{qh}(\mathbf{r},t)=\sum_{i=1}^{N_{qh}} V_\circ\,\delta^{(2)}(\mathbf{r}-\mathbf{r}^{({\rm lab})}_i(t))$ and will serve to create quasi-holes in the photon gas. Among the many techniques that are available to exert a potential on a photon gas~\cite{ICCC_RMP}, the all-optical techniques demonstrated in~\cite{Amo_potential,hayat} appear most promising for our purpose, as they combine a relatively strong potential with the fast modulation speed {needed} to braid the quasi-holes around.

{Photon-photon interactions are modeled at the simplest level via a contact repulsive potential of strength $g_{nl}$ proportional to the frequency-independent $\chi^{(3)}$ nonlinearity of the cavity medium.
A most promising choice to maximize the strength of interactions is to use an excitonic optical transition in  a solid-state quantum well~\cite{ICCC_RMP} for which reinforced nonlinearities via the biexciton Feshbach mechanism~\cite{feshbach,Deveaud_private} or via mixing with an indirect exciton~\cite{Cristofolini} are presently under active investigation. Another possible option is to employ a cloud of optically dressed atoms in a Rydberg EIT configuration for which strong nonlinearities have been recently observed~\cite{Peyronel}.}
As we shall see in the following, efficient generation of a Laughlin state of photons requires that the interaction-induced gap exceeds the dissipation-induced broadening of the states.

Injection of photons into the cavity by a coherent pump laser is described by the last line of the Hamiltonian, the spatio-temporal profile of the laser being fixed by the function $F(\mathbf{r},t)$: in the following, we shall restrict our attention to the case of a monochromatic pump {of} frequency $\omega_p^{({\rm lab})}$ and {normalized} amplitude $F$ with the spatial profile of a Laguerre-Gauss beam ${\rm LG}_0^m$ centered on the cavity axis with orbital angular momentum $m$ (in units of $\hbar$).
As it happens in any optical system, photons in the cavity have a finite lifetime and eventually decay at a rate $\gamma$. This can be taken into account in our theoretical model by including Lindblad terms in the master equation for the density matrix~\cite{Walls}. As a result, the steady-state of the photon gas will be determined by a dynamical balance of pumping and losses. In particular, the overall rotation of the cloud is continuously supported by the angular momentum that is injected into the cavity by the coherent pump.

In the following, it will be useful to describe the system from a reference frame rotating at angular frequency $\Omega$ around $\mathbf{\hat{z}}$.  To this purpose, one can either add to the Hamiltonian (\ref{eq:H}) a single term proportional to the total angular momentum $L_z$ along the rotation axis $\hat{\mathbf{z}}$, $\mathcal{H}_\Omega = \mathcal{H}-\Omega L_z$, or include the centrifugal force as a reduction of the effective trapping frequency $\omega^2 \rightarrow \omega^2-\Omega^2$ and then separately account for the Coriolis force in terms of a vector potential $\mathbf{A}(\mathbf{r})=m_{ph}\, \Omega\,\hat{\mathbf{z}} \times \mathbf{r}$ minimally coupled to the photon momentum as $-i\hbar\nabla \rightarrow -i\hbar\nabla-\mathbf{A}(\mathbf{r})$.
While the former formulation is most convenient in calculations, {the} latter one emphasizes the close analogy with the dynamics of a charged particle in a magnetic field. Of course, in the rotating reference frame, the positions $\mathbf{r}_i$ of the delta potentials have to be accordingly rotated back by an angle $\Omega t$ with respect to the laboratory frame ones $\mathbf{r}_i^{({\rm lab})}$, and the pump frequency is shifted to $\omega_{p}=\omega_p^{({\rm lab})}- m \Omega$. In the following, pump frequencies will be measured from the photon rest frequency as $\Delta \omega_p=\omega_p-\omega_{c}$.

\section{Theory and Results}
\subsection{Laughlin state of photons}

Based on the Hamiltonian (\ref{eq:H}), we now discuss how it is possible to generate a Laughlin state of photons in the cavity without repulsive delta potentials ($V_\circ=0$). We start by considering the isolated system Hamiltonian $\mathcal{H}_\Omega$ in the frame rotating at $\Omega$ in the absence of driving ($F = 0$) and losses ($\gamma = 0$). When $\Omega\to \omega$, this Hamiltonian is seen to be formally identical to the one describing the FQH physics of interacting electrons in a magnetic field, if one replaces the Coulomb interactions with the present contact interactions. In particular, the exact $N$-particle ground state of $\mathcal{H}_\Omega$ is represented by the $\nu = 1/2$ bosonic Laughlin wave function
\be\Psi_{\rm FQH}(z_1, \ldots, z_N) = \prod_{j<k}(z_j-z_k)^2e^{-\sum_{l = 1}^N|z_l|^2/2},\label{WF_FQH}\ee
where $z_j = (x_j+iy_j)/\ell$ are the complex particle coordinates in units of the oscillator length $\ell=\sqrt{\hbar/m_{ph}\omega}$ \cite{Laughlin, Zoller_anyon}.
This wave function is composed of lowest Landau level (LLL) wave functions, has a total angular momentum $M = N(N-1)$ and is separated from the excited states by an energy gap approximately given by $g_{nl}/4\pi\ell^2$ in the low $g_{nl}$ limit where the LLL approximation is valid (cfr. Sec. I of the Supplemental Material).

It should be stressed from the outset that some properties of the Laughlin state that are connected to the fixed filling fraction $\nu = 1/2$, like the incompressibility and a constant anyonic braiding phase, will get finite-size corrections in the harmonic trap geometry under consideration, in particular when $N$ is not macroscopic~\cite{Fetter,Cooper_review}. However, as the wave function (\ref{WF_FQH}) represents the exact and unique ground state in the presence of contact interactions, regardless of $N$, one can unambiguously address it by optical means and still extract non-trivial information by studying its properties for the few-particle case.

Inspired from our previous works \cite{IC_TG,FQH_photon}, we propose to take advantage of the driven-dissipative nature of the photonic system to create such a Laughlin state in an all-optical way by shining onto the cavity a coherent pump with a Laguerre-Gauss ${\rm LG}_0^m$ transverse profile: given the cylindrical symmetry of our set-up, the orbital angular momentum $m$ has to match the value $N-1$ of the angular momentum per particle of the target $N$-particle Laughlin state. The efficiency of this strategy is explored by means of Monte Carlo wave function calculations of the steady-state density matrix under the combined effect of continuous-wave pumping and losses~\cite{FQH_photon,MCWF}. The results for the simplest $N = 2,3$ cases are summarized in Fig. \ref{Fig1}(b,c).

As we are using a coherent laser pump, the system is driven into a superposition of states with different number of particles. {However, several mechanisms can be exploited to efficiently isolate the contribution of states with the given $N$ of interest.}
On one hand, the contribution of all states with $N^{\prime}<N$ is eliminated by looking at the probability $P_N$ of having $N$ photons in the cavity, a quantity that can be extracted from a coincidence measurement of $N$ transmitted photons. In order to isolate the final $N$-photon resonance from the spectral features due to the intermediate states with $N^{\prime}<N$ photons, it is enough to plot the relative probability $P_N/P_{N-1}$ as a function of pump frequency $\Delta\omega_p$~\cite{FQH_photon}.

On the other hand, the excitation probability of higher-$N$ states can be strongly suppressed just by working in the weak driving limit $\ell F/\gamma\ll 1$ where the population of the $N^{\prime}$ photon state scales as $(\ell F/\gamma)^{2N^{\prime}}$. A further suppression of $N^{\prime}>N$ states is provided by a sort of {\em quantum Hall blockade effect} due to the quantum Hall gap: as the angular momentum per particle in the $N^{\prime}$-particle Laughlin state is larger than the angular momentum $N-1$ per injected photon, all accessible $N^\prime$-particle states lie above the quantum Hall gap and therefore cannot be resonantly excited by the coherent pump.

In Fig. \ref{Fig1}(b), we show a simulated spectrum of this quantity in the frame rotating at $\Omega=\omega$. For sufficiently low photon losses, sharp resonance peaks corresponding to the $N$-particle eigenstates of the isolated system appear in the spectrum. As originally
discussed in~\cite{IC_TG}, the position of the transmission peak is related to the $N$-body eigenenergies by the resonance condition $\omega_p = \omega^{(N)}/N=\omega_c+\omega$. For both $N=2,3$, the main peaks at $\Delta\omega_p/\omega=1$ correspond to an $N$-photon transition from vacuum to the lowest $N$-particle eigenstate of $H_\Omega$ at energy $\hbar \omega^{(N)} = N\hbar(\omega_c+\omega)$ fixed by the zero-point motion in the harmonic potential, which suggests that particles are non-overlapping in this state. For sufficiently strong interactions $g_{nl}/4\pi N\ell^2 \gg \gamma$, this peak is well separated from the ones corresponding to excited states within the same $N$-photon manifold.

As a {further} check of the Laughlin nature of the {generated $N$-particle} state, we can look at the overlap $\mathcal{O}(\Psi_{\rm FQH},\Phi)=|\langle\Psi_{\rm FQH}|\Phi\rangle|^2/\langle\Psi_{\rm FQH}|\Psi_{\rm FQH}\rangle\langle\Phi|\Phi\rangle$ between the $N$-photon amplitude $\Phi(z_1,\ldots,z_N) = {\rm Tr}[\rho_{ss}\hat{\Psi}(z_1)\ldots\hat{\Psi}(z_N)]$ and the target Laughlin wave function $\Psi_{\rm FQH}$. As we discussed in~\cite{FQH_photon}, in the weak driving limit, the $N$-photon amplitude gives in fact the many-body wave function of the single $N$-particle pure state reached by the system and is experimentally accessible from repeated measurements of the field quadratures of the transmitted light. Its dependence on the pump frequency $\Delta\omega_p$ is shown in Fig. \ref{Fig1}(c): as expected, the maximum overlap is obtained at $\Delta\omega_p/\omega=1$; its peak value larger than $99.5\%$ confirms that the generated state is basically the $N$-particle Laughlin state. As angular momentum
of the
rotating gas
is continuously replenished by the pump beam, the photon system is much less sensitive to trap anisotropies than atomic clouds~\cite{Fetter,Cooper_review,Dalib_rev}: as a result, the overlap with the Laughlin state is still $\approx 97\%$ for trap anisotropies as large as {$(\omega_x-\omega_y)/(\omega_x+\omega_y)=0.01$ (which corresponds to $\omega_x-\omega_y\approx \gamma$).}

\subsection{Quasi-hole braiding}

We can now turn to the generation of quasi-hole states in our system. This can be done by adding localized repulsive potentials to pierce holes in the photon gas. As it is sketched in Fig. \ref{Fig1}(a), their position $\mathbf{r}^{({\rm lab})}_i(t)$ in the laboratory frame is assumed to be rotating at an angular frequency $\Omega$ around the cavity axis, so as to be stationary at $\mathbf{r}_i$ in the frame rotating  at $\Omega$.
In the absence of pumping and losses, and for $\Omega\to \omega$, the ground state of the one-quasi-hole Hamiltonian $\mathcal{H}_\Omega^\circ$ is successfully represented by the single quasi-hole wave function \cite{Yoshioka,Laughlin,Zoller_anyon}.
\be \Psi_\circ(z_1,\ldots, z_N) = \prod_i(z_i-z_\circ)\,\Psi_{\rm FQH}(z_1,\ldots, z_N), \label{WF_1qh}\ee
where $z_\circ = r_\circ e^{i\theta_\circ}/\ell$ is the complex coordinate of the quasi-hole  (cfr. Sec. I of the Supplemental Material).
Another quasi-hole sitting, e.g., at the center of the trap $z_{\circ\circ}=0$ can be included via a second delta-function potential in the two quasi-hole Hamiltonian $\mathcal{H}_\Omega^{\circ\circ}$. Again, in the $\Omega \to \omega$ limit, the ground state wave function can be written in the simple form
\be \Psi_{\circ\circ}(z_1,\ldots, z_N) = \prod_i(z_i-z_\circ)z_i\,\Psi_{\rm FQH}(z_1,\ldots, z_N).
\label{WF_2qh}
\ee

The crucial point of our proposal is to relate the braiding phase observed in the reference frame rotating at the trap frequency $\omega$ to the time-independent energy spectrum in the frame rotating at slightly lower $\Omega=\omega-\delta\Omega$. In the frame rotating at $\omega$, the quasi-hole at $z_\circ$ is in fact slowly rotating at frequency $\delta\Omega$ in the backwards direction: provided $\delta\Omega$ is small enough, this process is equivalent to adiabatically looping the quasi-hole at $z_\circ$ along the circle of radius $r_\circ$ following the position of the localized potential. As a result, after a rotation period $T = 2\pi/\delta\Omega$, the quasi-hole will return to its original position~\cite{Berry}, with the single (double) quasi-hole wave function $\Psi_\circ$ ($\Psi_{\circ\circ}$) having acquired a Berry phase $\phi_B^\circ$ ($\phi_B^{\circ\circ}$) in addition to the trivial dynamical phase $E_\omega^{\circ} T$ ($E_\omega^{\circ\circ} T$).

When observed from the reference frame rotating at $\Omega$ where the localized potentials are fixed in space, the time evolution reduces for any $t$ to the phase $E^{\circ,\circ\circ}_\Omega t$. At time $t=T$ when a rotation is complete, the wave functions in the two frames have to coincide again, which establishes a relation between the energy difference $\Delta E^{\circ,\circ\circ}=E^{\circ,\circ\circ}_\omega - E^{\circ,\circ\circ}_\Omega$ and the many-body Berry phases $\phi^{\circ,\circ\circ}_B$,
\be \phi^{\circ,\circ\circ}_B = 2\pi\frac{\Delta E^{\circ,\circ\circ}}{\hbar\,\delta\Omega} \:\:\:({\rm mod}\:2\pi).
\label{Berry_Energyshift}\ee
This relation holds for both quasi-hole states in a quantum Hall liquid, as well as in a non-interacting system (cfr. Sec. III of the Supplemental Material). As it relates the many-body Berry phase to spectroscopically observable quantities such as the energies, it will be the basis  of the measurement scheme we are now going to illustrate.

\begin{figure}[h]
\includegraphics[width = \columnwidth,clip]{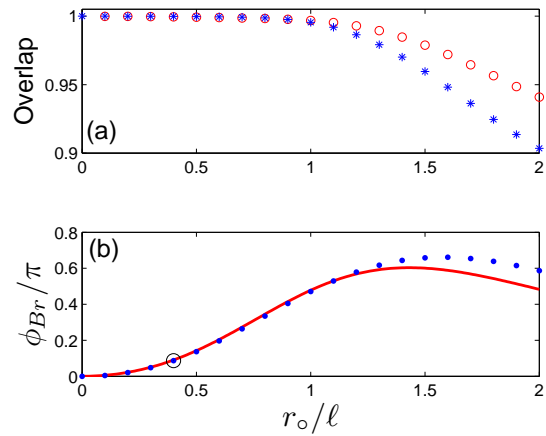}
\caption{(a) Stars $\ast$ (circles $\circ$) show the overlap of the $N=2$ lowest energy eigenstate of $\mathcal{H}_\Omega^{\circ}$ ($\mathcal{H}_\Omega^{\circ\circ}$) with the one (two) quasi-hole wave function $\Psi_\circ$ ($\Psi_{\circ\circ}$) as a function of the position $r_\circ$ of the exterior quasi-hole.
(b) Estimation ($\bullet$) of the braiding phase calculated via Eq. (\ref{Berry_Energyshift}) compared with the analytical result from the quasi-hole wave functions $\Psi_\circ$, $\Psi_{\circ\circ}$ (solid line).
System parameters $g_{nl}/\ell^2\omega=4$, $\Omega/\omega = 0.99$ and $V_\circ/\ell^2\,\hbar\omega = 100$.\label{Fig2} }
\end{figure}

\subsection{Numerical results and discussion}

As a first step, we wish to numerically confirm the validity of Eq. (\ref{Berry_Energyshift}) for the isolated system. To this purpose, we look for the ground state wave functions in the rotating frame at $\Omega$ where the quasi-holes are fixed in space and the Hamiltonians $\mathcal{H}_\Omega^{\circ}$ and $\mathcal{H}_\Omega^{\circ\circ}$ are time-independent. Their overlap with the analytic wave-functions (\ref{WF_1qh}) and (\ref{WF_2qh}) for $\Omega/\omega=0.99$ is shown in Fig. \ref{Fig2}(a) as a function of the position $r_\circ$ of the exterior quasi-hole.
In the lower panel, we show the value of the braiding phase $\phi_{\rm Br} = \phi_B^\circ-\phi_B^{\circ\circ}$: as introduced in~\cite{Wilczek_anyon}, this is the difference between the many-body Berry phases acquired by the ground state wave functions in the presence of single and double delta-function potentials. In this panel, the value of $\phi_{\rm Br}$ extracted via Eq. (\ref{Berry_Energyshift}) from experimentally accessible quantities is compared with the result of a direct calculation of the Berry phases from the analytical wave functions (\ref{WF_1qh}) and (\ref{WF_2qh}) (cfr. Sec. II of the Supplemental Material). The agreement is excellent up to a radius $r_\circ \approx \ell$, i.e. when the quasi-hole potential starts exiting the cloud: at this point, the repulsive delta potential is no longer able to sustain the quasi-hole state and the overlap shown in panel (a) suddenly drops.

\begin{figure}[h]
\includegraphics[width = \columnwidth,clip]{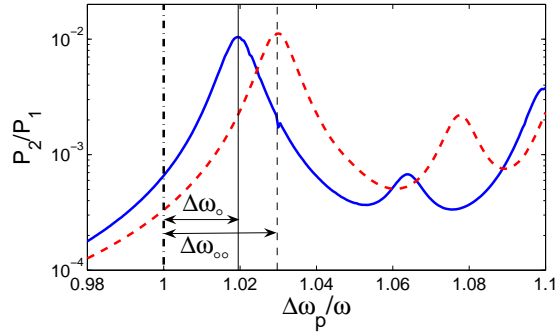}
\caption{Steady-state relative probability $P_2/P_1$ of two-particle excitation for $\mathcal{H}_\Omega^{\circ}$ with a Laguerre-Gauss ${\rm LG}_0^2$ pump (solid line), and for $\mathcal{H}_\Omega^{\circ\circ}$ and an ${\rm LG}_0^3$ pump (dashed line). Vertical solid (dashed) lines correspond to half the eigenenergies of the isolated system. Shift of the lowest transmission peak with respect to $\Delta\omega_p/\omega=1$ is denoted by $\Delta\omega_{\circ,\circ\circ} \equiv \Delta E^{\circ,\circ\circ}/2\hbar\omega$. System and pump parameters as in Fig. \ref{Fig1}, with $\Omega/\omega=0.99$, $r_{\circ}=0.4$ and $V_\circ/\ell^2\hbar \omega = 100$.~\label{Fig3}}
\end{figure}

The optical generation of the quasi-hole states is then illustrated in Fig. \ref{Fig3}: a monochromatic Laguerre-Gauss beam is shined on the cavity in the presence of the repulsive potentials rotating at a frequency $\Omega/\omega=0.99$. An efficient choice for the orbital angular momentum of the pump is to use the closest integer to the angular momentum per particle of the target state. The steady-state density matrix is numerically calculated via Monte Carlo wave function technique in the frame rotating at $\Omega$ where the Hamiltonian is time-independent. The different curves in the figure show the spectrum of $P_2/P_1$ as a function of pump frequency $\Delta\omega_p$: the solid (dashed) curve refers to the one (two) quasi-hole Hamiltonian $\mathcal{H}_\Omega^{\circ}$ ($\mathcal{H}_\Omega^{\circ\circ}$) including one (two) localized repulsive potential.
The most relevant feature in these spectra is the well isolated lowest frequency peak at $\Delta\omega_p/\omega=1.0195$ on the solid line and $1.0297$ on the dashed one: their identification with quasi-hole states is confirmed by the excellent overlap $\approx 99\%$ of the two-photon amplitude with the analytical wave functions in (\ref{WF_1qh}) and (\ref{WF_2qh}).
Remembering that these two-photon peaks are located at half the energy $E_\Omega^{\circ,\circ\circ}$ of the two-photon eigenstate, it is then straightforward to extract via (\ref{Berry_Energyshift}) the value of the many-body Berry phase when one quasi-hole at $r_\circ$ is braided around another one located at the center of the trap: as one can see from the small circle in Fig. \ref{Fig2}(b), the accuracy of this simulated measurement is excellent. {This result is confirmed by an analogous calculation of the position of three-photon peaks performed within the LLL approximation (cfr. Sec. IV of the Supplemental Material). In spite of the obvious technical difficulties, we expect that the proposed protocol to measure the many-body Berry phase should be applicable also to states with a macroscopic number of photons for which the theory of the fractional quantum Hall effect would predict an anyonic braiding phase of $\pi$~\cite{Yoshioka,Wilczek_anyon,Lewenstein}.}

\section{Conclusions}

In conclusion, we have proposed and characterized an all-optical scheme to generate and manipulate few-particle quantum Hall states of strongly interacting photons in a nonlinear optical cavity. Quasi-holes in the photon gas can be pierced and braided with repulsive potentials and the corresponding many-body Berry phase can be detected from the spectral shifts of the resonant transmission peaks. Extension of this work to more complex configurations involving e.g. light polarization degrees of freedom may open the way to observe anyonic excitations with non-Abelian statistics.

\section{Acknowledgements}
We are grateful to A. Imamo\u glu, R. Santachiara, {T. Volz and M. Fleischhauer's group for stimulating exchanges. This work has been supported by ERC through the QGBE grant and by Provincia Autonoma di Trento}.

\end{document}


\title{Supplemental Material}

\maketitle

\section{Dependence of eigenstates and eigenenergies on the relative angular frequency $\delta\Omega$}

In this section we justify the adiabatic following of the ground state when the quasi-hole potential is braided around in space.

\begin{figure}[htbp]
\includegraphics[width = \columnwidth,clip]{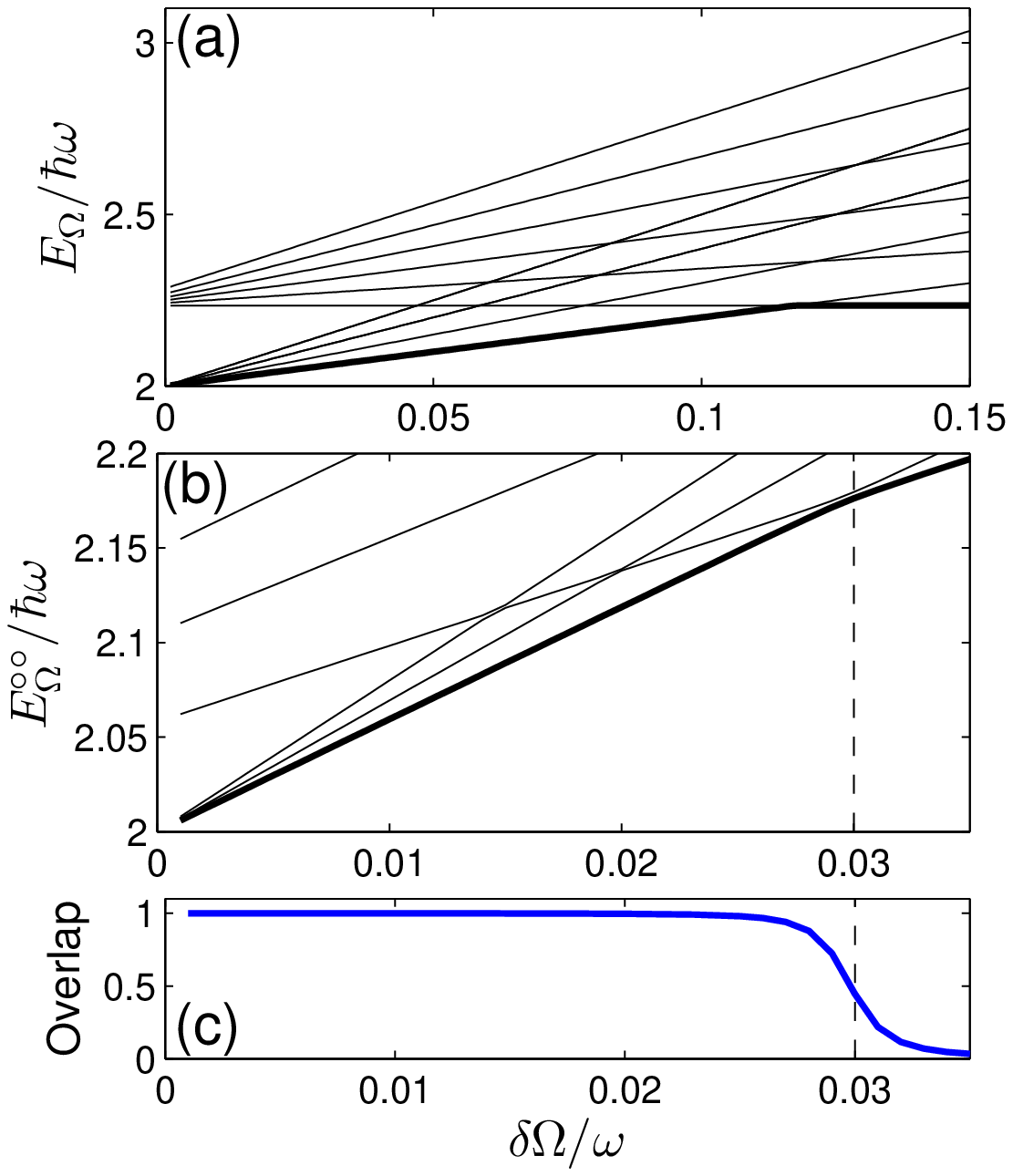}
\caption{(a) Energy of the lowest $N=2$ particle eigenstates of $\mathcal{H}_\Omega$ as a function of relative rotation frequency $\delta\Omega=\omega-\Omega$ in the absence of the repulsive quasi-hole potentials. For clarity, only the states of total angular momentum $L_z\leq 5$ are displayed.
(b) Energy of the lowest $N=2$ particle eigenstates of the two quasi-hole Hamiltonian $\mathcal{H}^{\circ\circ}_\Omega$ as a function of relative rotation frequency $\delta\Omega=\omega-\Omega$. For clarity, only states at low average angular momentum $\langle L_z \rangle < 8$ are displayed.
(c) Overlap of the ground state of $\mathcal{H}^{\circ\circ}_\Omega$ with the analytical two quasi-hole wave function, Eq. (4) of the main text.
System parameters: $g_{nl}/\ell^2\omega=4$, $V_\circ/\ell^2\,\hbar\omega = 100$. The exterior quasi-hole is positioned at $r_{\circ}=0.4\ell$. \label{FigS1}}
\end{figure}

Let us start from the Hamiltonian $\mathcal{H}_\Omega$ with no repulsive localized potentials. The energies of the eigenstates of $\mathcal{H}_\Omega$ are plotted in Fig. \ref{FigS1}(a) as a function of $\delta \Omega$.
For $\delta\Omega\to 0$, we have a sequence of massively degenerate states: the lowest manifold corresponds to wave functions obtained as the product of a Laughlin state $\Psi_{FQH}$, as defined in Eq. (3) of the main text, times any polynomial symmetric under $z_1 \leftrightarrow z_2$ and is separated from the higher manifold by a gap mostly determined by the interaction strength $g_{nl}$. Within the lowest Landau Level (LLL) approximation valid for small $g_{nl}/\ell^2 \ll\hbar \omega$, the gap is proportional to $g_{nl}$ and close to the integral $g_{nl}\int|\psi_0(\mathbf{r})|^4d^2\mathbf{r}/2 =g_{nl}/4\pi\ell^2$, $\psi_0$ being the harmonic oscillator ground state wave function.
For finite $\delta \Omega$, the states of the first manifold are split according to their total angular momentum. The non-degenerate ground state has the lowest angular momentum $\hbar N(N-1)$ and is given by the Laughlin state $\Psi_{FQH}$. The first excited state has one more unit of angular momentum and is separated from the ground state by $\hbar\,\delta\Omega$.

We have then performed a numerical diagonalization of the Hamiltonian $\mathcal{H}_\Omega^{\circ,\circ\circ}$ in the presence of one and two localized repulsive potentials piercing quasi-holes through the quantum Hall liquid. The result for the two quasi-hole case is shown in Fig. \ref{FigS1}(b): for small but non-zero $\delta\Omega$, the ground state is separated from the first excited state by a small amount close to $\hbar\,\delta\Omega$. The next manifold lies higher in energy by an amount determined by $g_{nl}$. States involving non-lowest-Landau-level components are found at even higher energies of the order of $\hbar\omega$.
The quasi-hole nature of the ground state is numerically confirmed by a calculation of its overlap with the two quasi-hole wave function, as given by Eq. (4) of the main text. The higher states of the manifold correspond to wave functions obtained by multiplying this two quasi-hole wave function by arbitrary symmetric polynomials.

The fact that the two quasi-hole state is the ground state even for non-zero values of $\delta\Omega$ confirms that holes are indeed braided around the quantum Hall fluid without creating excitations provided braiding is performed slowly enough. As we have seen in the main text, the overlap with the quasi-hole wave function suddenly drops at some value of the quasi-hole position $r_\circ$. An analogous phenomenon is visible here as a function of angular frequency $\delta\Omega$ in Fig. \ref{FigS1}(c): overlap suddenly drops at the first avoided crossing point between the ground state and a lower angular momentum state from the higher manifold.

\section{Berry Phase for one and two quasi-hole states}

Here we prove a simple relation between the expected value of total angular momentum $\langle L_z\rangle$ and the Berry phase $\phi_{B}$ acquired by a many-body wave function lying in the LLL, when a quasi-hole is looped along a circle with or without a second quasi-hole sitting at the center of the circle.

The generic form of the $N$-particle wave function we consider is
\begin{multline}
\Psi(z_1,\ldots,z_N;z_\circ)= \sum_{n_1,\ldots,n_N} c_{n_1,\ldots,n_N}\,z_1^{n_1}\ldots z_N^{n_N} \\
\times(z_1-z_\circ)\ldots(z_N-z_\circ)\,e^{-\sum_{j=1}^N|z_j|^2/2},
\end{multline}
where $z_j = (x_j+iy_j)/\ell$ are the complex particle coordinates, $z_\circ$ is the quasi-hole coordinate and $n_j$ are nonnegative integers.
The only assumption we are making is that $\sum_{j=1}^Nn_j=A$ is constant, meaning that the part of the wave function shown in the first line is a homogenous multivariate polynomial of degree $A$. A second quasi-hole term $z_1\ldots z_N$ can be included into this part by letting each $n_j$ increase by $1$.

For simplicity, from now on we will omit displaying the exponential factor which is common to all terms in the summation over $n_j$ and use the shorthand notations $\{z\}\equiv z_1,\ldots,z_N$, and $\{n\}\equiv n_1,\ldots,n_N$. We will also expand the product $(z_1-z_\circ)\ldots(z_N-z_\circ)$ as follows $(z_1-z_\circ)\ldots(z_N-z_\circ) = \sum_{l=0}^N d_l z_\circ^l $. Upon these changes, the wave function becomes:
\be \Psi(\{z\};z_\circ) \equiv \sum_{\{n\}}c_{\{n\}}z_1^{n_1}\ldots z_N^{n_N}\sum_{l=0}^N d_l z_\circ^l.\label{WF}\ee
Let us now calculate $\phi_{B}$ for a quasi-hole at $z_\circ = r_\circ e^{i\theta_\circ}/\ell$ with $\bar{r}_\circ\equiv r_\circ/\ell$ fixed:
\begin{multline}
 \phi_{B}= -i\int_0^{2\pi}\!\!\!\int d\{z\}\,d\{z^{\ast}\}\,\Psi^{\ast}\,\partial_{\theta_\circ}\Psi\,d\theta_\circ= \\
= 2\pi\!\!\int\!\! d\{z\}\,d\{z^{\ast}\}\!\!\!\sum_{\{n\},\{n^{\prime}\}}\!\!c^{\ast}_{\{n^{\prime}\}}c_{\{n\}}z_1^{\ast n_1^{\prime}}z_1^{n_1}
\ldots z_N^{\ast n_N^{\prime}}z_N^{n_N}\\ \times \sum_{l=0}^N l |d_l|^2\bar{r}_\circ^{2l},\label{Berry}
\end{multline}
where $d\{z\}\equiv dz_1\ldots dz_N$ and $\partial_{\theta_\circ} \equiv \partial/\partial\theta_\circ$ is the partial derivative with respect to $\theta_\circ$.
Now let us look at the expected value of total angular momentum:
\bea
\!\!\!\langle L_z\rangle\!\! &=& \!\!\int\!\! d\{z\}d\{z^{\ast}\}\!\! \sum_{\{n^{\prime}\}} c^{\ast}_{\{n^{\prime}\}}z_1^{\ast n_1^{\prime}}\!\!\ldots z_N^{\ast n^{\prime}_N}\sum_{l^{\prime}=0}^N d^{\ast}_{l^{\prime}} z_\circ^{\ast l^{\prime}}\nonumber \\
&\times& \!\!\!(\!-i\hbar)(\partial_{\theta_1}\!\!+\ldots+\!\partial_{\theta_N})\!\!\sum_{\{n\}}c_{\{n\}}z_1^{n_1}\!\!\ldots z_N^{n_N}\!\!\sum_{l=0}^N d_l z_\circ^l.\label{L_z}
\eea
While taking the derivatives $\partial_{\theta_j}$ with respect to the angular variables (we have set $z_j=|z_j|\,e^{i\theta_j}$), we can use the chain rule to isolate the contribution of the $\sum_{l=0}^N d_l z_\circ^l$ term. In this way, the second line of Eq. (\ref{L_z}) becomes
\bea \hbar\sum_{\{n\}}c_{\{n\}}(n_1+\ldots + n_N)z_1^{n_1}\!\!\ldots z_N^{n_N}\!\!\sum_{l=0}^N d_l z_\circ^l \nonumber \\
+\hbar\sum_{\{n\}}c_{\{n\}}z_1^{n_1}\!\!\ldots z_N^{n_N}\!\!\sum_{l=0}^N (N-l)d_l z_\circ^l.\label{second line}\eea
To get this result one simply has to recall that $d_l = z_1^{s_1}\ldots z_N^{s_N}$ with $(s_1+\ldots+s_N) = N-l$. Inserting (\ref{second line})   into Eq. (\ref{L_z}), and reordering terms we obtain
\bea \langle L_z\rangle &=& \hbar\sum_{\substack{\{n\},\{n^{\prime}\}\\l,l^{\prime}}}c^{\ast}_{\{n^{\prime}\}}c_{\{n\}}z_\circ^{\ast l^{\prime}}z_\circ^l(A+N-l) \nonumber \\
&\times& \int d\{z\}d\{z^{\ast}\}z_1^{\ast n_1^{\prime}}\!\!\ldots z_N^{\ast n^{\prime}_N}d^{\ast}_{l^{\prime}}z_1^{n_1}\!\!\ldots z_N^{n_N}d_l. \eea
In order for the integral not to vanish, the total power of conjugated coordinates should match that of unconjugated ones: $A+N-l^{\prime} = A+N-l$, implying $l^{\prime}=l$. As a result, one gets
\begin{widetext}
\bea
-2\pi\frac{\langle L_z\rangle}{\hbar} &=& -2\pi(A+N)\!\!\!\!\sum_{\{n\},\{n^{\prime}\}}\!\!\!\!c^{\ast}_{\{n^{\prime}\}}c_{\{n\}}\!\!\int\!\! d\{z\}d\{z^{\ast}\}z_1^{\ast n_1^{\prime}}\!\!\ldots z_N^{n_N}\sum_{l=0}^N |d_l|^2\bar{r}_\circ^{2l}\nonumber\\
&+&2\pi\!\!\!\!\sum_{\{n\},\{n^{\prime}\}}\!\!\!\!c^{\ast}_{\{n^{\prime}\}}c_{\{n\}}\!\!\int\!\! d\{z\}d\{z^{\ast}\}z_1^{\ast n_1^{\prime}}\!\!\ldots z_N^{n_N}\sum_{l=0}^N l|d_l|^2\bar{r}_\circ^{2l}.
\eea
\end{widetext}
The first term apart from the factor $-2\pi(A+N)$ is the normalization of the wave function (\ref{WF}), which we take to be 1. The second term is $\phi_B$ we have found in Eq. (\ref{Berry}). Thus, finally we have
\be
\phi_{B} = -2\pi\frac{\langle L_z\rangle}{\hbar}+2\pi(A+N).
\label{Berry_AngMom}
\ee
This result can be explicitly checked for the simplest $N=1$ and $N=2$ cases, for which one can analytically calculate $\phi_B$ and $\langle L_z\rangle$ without much difficulty. Here we do not give the details of the elementary verification and only quote some final results for one and two quasi-hole states.

For the case of $N=1$, the Berry phase for the single-quasi-hole wave function $\propto(z_1-z_\circ)e^{-|z_1|^2/2}$ (here $A=0$) is
\be\phi_{B}^{1,\circ} = \frac{2\pi \bar{r}_\circ^2}{1+\bar{r}_\circ^2},\label{Berry1p1qh}\ee
while for the double-quasi-hole wave function $\propto z_1(z_1-z_\circ)e^{-|z_1|^2/2}$ (here $A=1$) it is
\be\phi_{B}^{1,\circ\circ} = \frac{2\pi \bar{r}_\circ^2}{2+\bar{r}_\circ^2}.\label{Berry1p2qh}\ee
For the case of $N=2$, the Berry phase for the $\nu=1/2$ Laughlin wave function with a single quasi-hole $\propto(z_1-z_2)^2(z_1-z_\circ)(z_2-z_\circ)e^{-(|z_1|^2+|z_2|^2)/2}$ (here $A=2$) is
\be\phi_{B}^{2,\circ} = 2\pi\frac{4\bar{r}_\circ^2+4\bar{r}_\circ^4}{7+4\bar{r}_\circ^2+2\bar{r}_\circ^4},\label{Berry2p1qh}\ee
while for the double-quasi-hole wave function $\propto(z_1-z_2)^2z_1z_2(z_1-z_\circ)(z_2-z_\circ)e^{-(|z_1|^2+|z_2|^2)/2}$ (here $A=4$) it is
\be\phi_{B}^{2,\circ\circ} = 2\pi\frac{18\bar{r}_\circ^2+14\bar{r}_\circ^4}{60+18\bar{r}_\circ^2+7\bar{r}_\circ^4}.\label{Berry2p2qh}
\ee
To summarize, the relation given in Eq. (\ref{Berry_AngMom}) provides a direct way to efficiently calculate the Berry phase as soon as the coefficients of each term in the wave function are known \cite{NACLAB}: in the last section we shall see an example of its application to states with a larger number of particles $N$ up to $6$.

In addition to this, this relation suggests an alternative experimental method to determine the Berry phase for setups where a precise measurement of total angular momentum might be feasible.

\section{Results for the non-interacting system}

\begin{figure}[htbp]
\includegraphics[width = \columnwidth,clip]{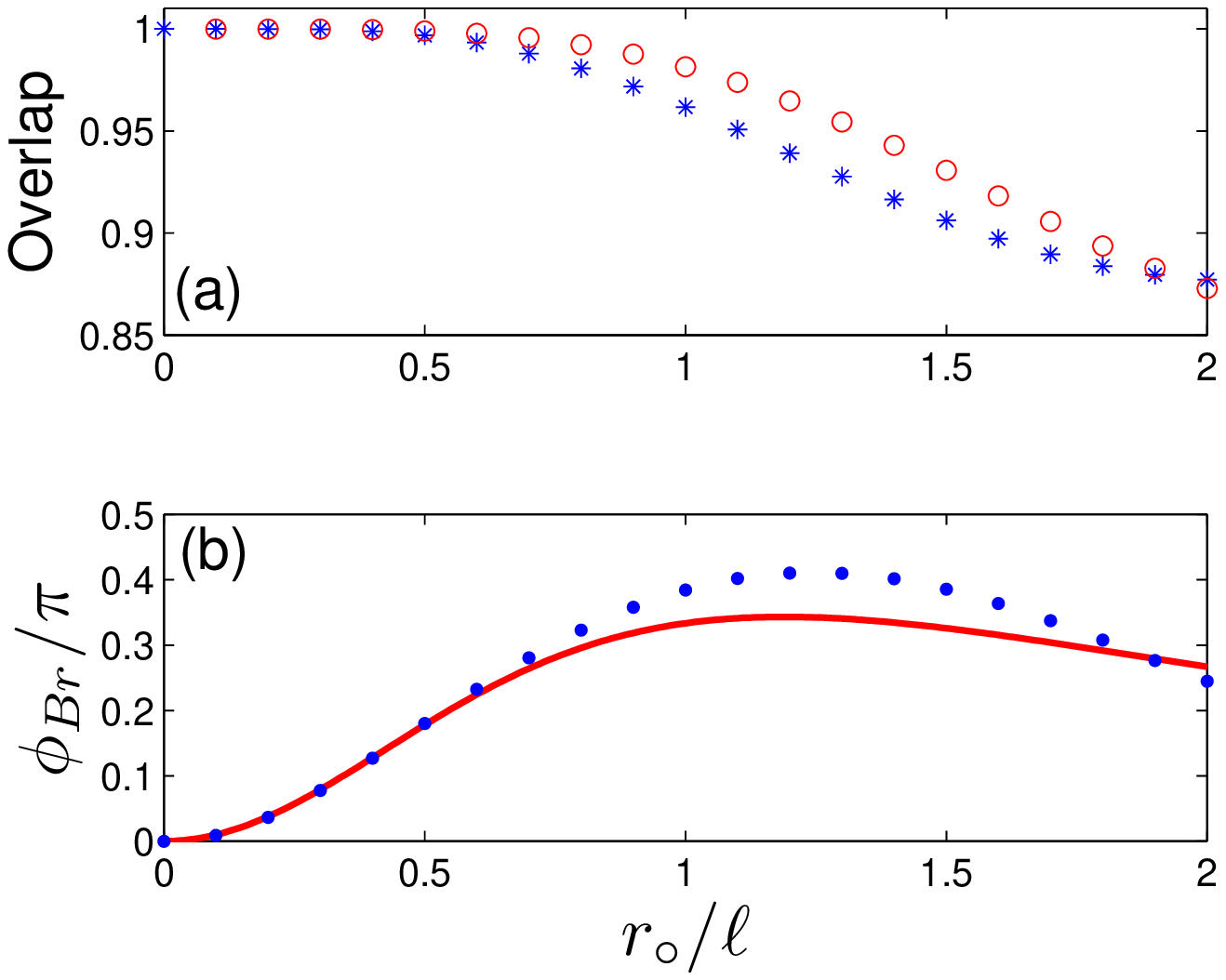}
\caption{(a) Stars $\ast$ (circles $\circ$) show the overlap of the lowest energy state wave function of a single particle in the presence of one (two) delta-function potential with the one (two) quasi-hole wave function as a function of the exterior quasi-hole coordinate $r_\circ$. System parameters: $\Omega/\omega = 0.99$, $V_\circ/\ell^2\hbar\omega = 100$. (b) Braiding phase calculated from Eq. (\ref{Berry_Energyshift}) is shown by dots ($\bullet$); the curve is the analytical result for $\phi_{B}^{1,\circ}-\phi_{B}^{1,\circ\circ}$ as given by Eqs. (\ref{Berry1p1qh}) and (\ref{Berry1p2qh}).\label{FigS2}}
\end{figure}

In order to get a deeper understanding of the braiding phase in a quantum Hall liquid, it is interesting to study the structure of the ground state and calculate the Berry phase also in the different case of a non-interacting system with $g_{nl}=0$.

With no loss of generality, we can start from the $N=1$ single particle case and calculate the braiding phase $\phi_{Br}$ in the presence of one and two delta-function potentials using the relation (5) of the main text
\be \phi^{\circ,\circ\circ}_B = 2\pi\frac{\Delta E^{\circ,\circ\circ}}{\hbar\,\delta\Omega} \:\:\:({\rm mod}\:2\pi).
\label{Berry_Energyshift}\ee
The many-body wave function for a larger number $N$ of non-interacting particles is in fact equal to the product of $N=1$ wave functions, so that both the energy and the braiding phase are just multiplied by a factor $N$.

The braiding phase as a function of the exterior quasi-hole  coordinate $r_\circ$ is plotted in Fig. \ref{FigS2}(a) and compared with the difference $\phi_{B}^{1,\circ}-\phi_{B}^{1,\circ\circ}$ calculated from the explicit expressions in Eqs. (\ref{Berry1p1qh}) and (\ref{Berry1p2qh}): once again, the agreement is excellent as long as the repulsive potential is well inside the cloud, $r_\circ\lesssim 0.5\ell$. This explanation is validated in panel (b) where we show the overlap of the ground state with one and two repulsive potentials with the corresponding analytical forms of the one and two quasi-hole wave functions.

\begin{figure}[htbp]
\includegraphics[width = \columnwidth,clip]{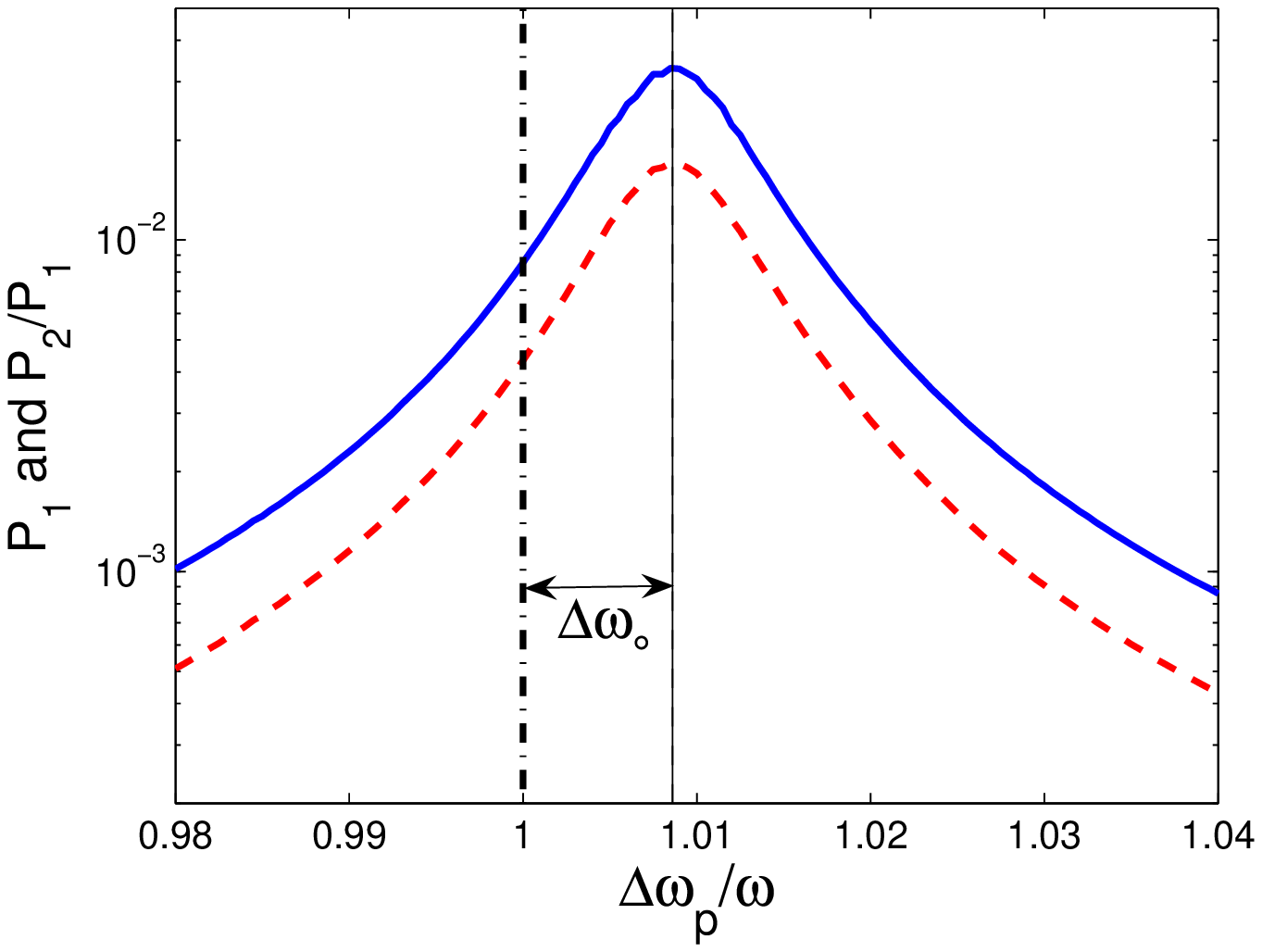}
\caption{Steady-state probability of single-particle excitation $P_1$ (solid line) and the relative probability of two-particle excitation $P_2/P_1$ (dashed line) in the presence of a single delta potential localized at $r_\circ=0.4\ell$ under a monochromatic Laguerre-Gauss ${\rm LG}_0^1$ pump. System and pump parameters: $\Omega/\omega = 0.99$, $\gamma/\omega = 0.01$, $\ell F/\gamma = 0.1$. \label{FigS3}}
\end{figure}

The optical response of the $g_{nl}=0$ system is a purely linear one: transmission spectra for a Laguerre-Gauss ${\rm LG}_0^1$  mode with one unit of angular momentum  are shown in Fig. \ref{FigS3} for the single quasi-hole case. As expected, the position of the one photon peak matches the ground state energy of the isolated system.
The two-photon peak in the $P_2/P_1$ spectrum occurs at the same frequency, as expected from the simple fact that the two-photon ground-state energy is twice the single-particle ground-state energy.
The Berry phase calculated from the shift $\Delta\omega_{\circ}$ of the peak via Eq. (\ref{Berry_Energyshift}) is $\approx0.280\pi$, in good agreement with the analytical prediction $\approx0.276\pi$ from Eq. (\ref{Berry1p1qh}).

\section{Results with more than two interacting particles}

\begin{figure}[htbp]
\includegraphics[width = \columnwidth,clip]{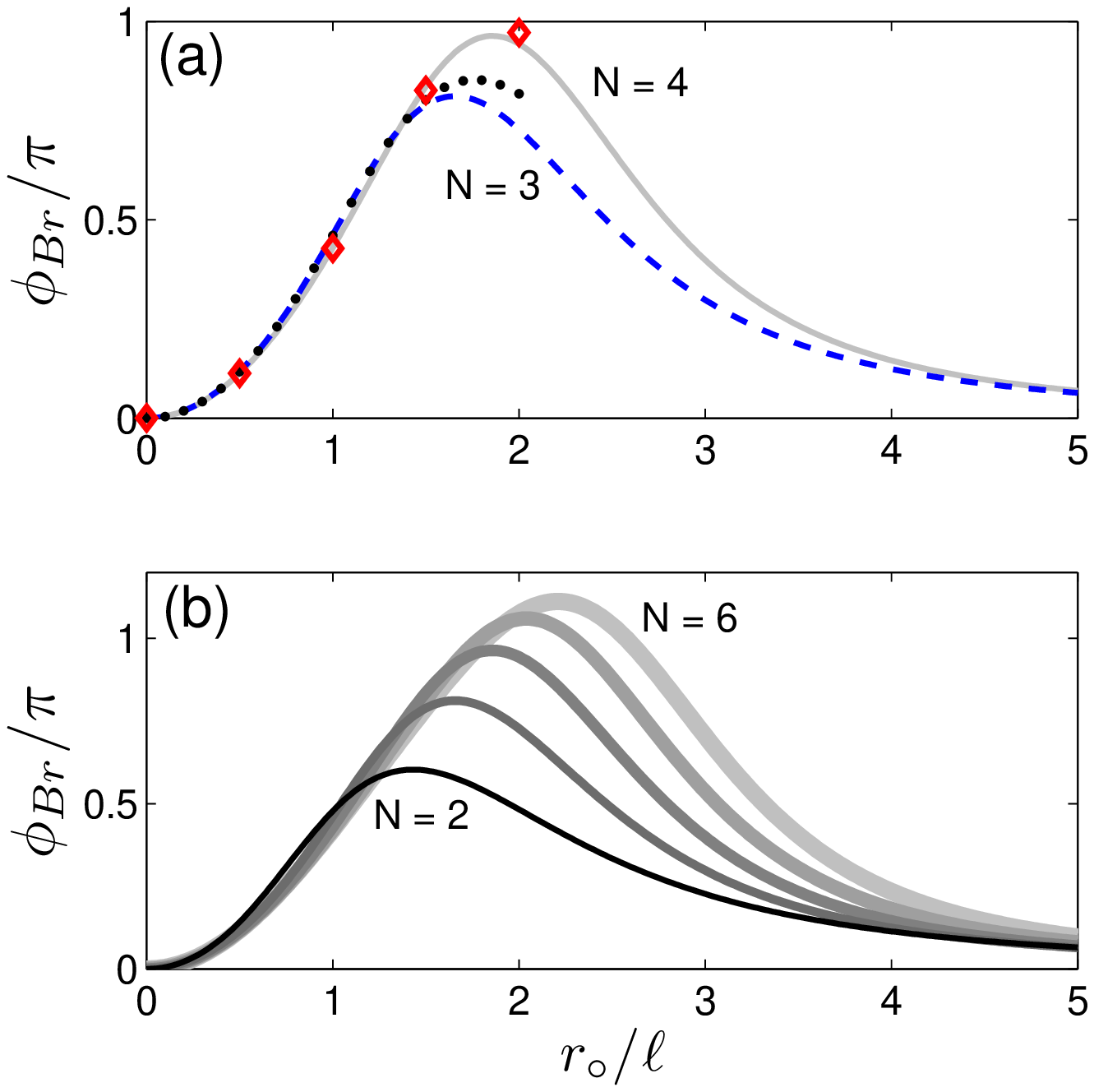}
\caption{(a) Braiding phase $\phi_{Br}$ calculated for $N=3$ (dashed line) and $N=4$ (solid line) from the analytical forms of the one and two quasi-hole wave functions, Eqs. (3) and (4) of the main text. Corresponding $\phi_{Br}$ extracted via Eq. (\ref{Berry_Energyshift}) from the numerical ground state of $\mathcal{H}_\Omega^{\circ,\circ\circ}$ within the LLL approximation for $N=3$ ($\bullet$) and $N=4$ ($\diamond$). System parameters: $\Omega/\omega = 0.99$, $g_{nl}/\ell^2\omega=4$, and $V_\circ/\ell^2\,\hbar\omega = 100$. (b) Braiding phase $\phi_{Br}$ calculated analytically for $N = 2,3,4,5$, and $6$ from the analytical forms of the one and two quasi-hole wave functions, Eqs. (3) and (4) of the main text. Line thickness increases with $N$.\label{FigS4}}
\end{figure}

\begin{figure}[htbp]
\includegraphics[width = \columnwidth,clip]{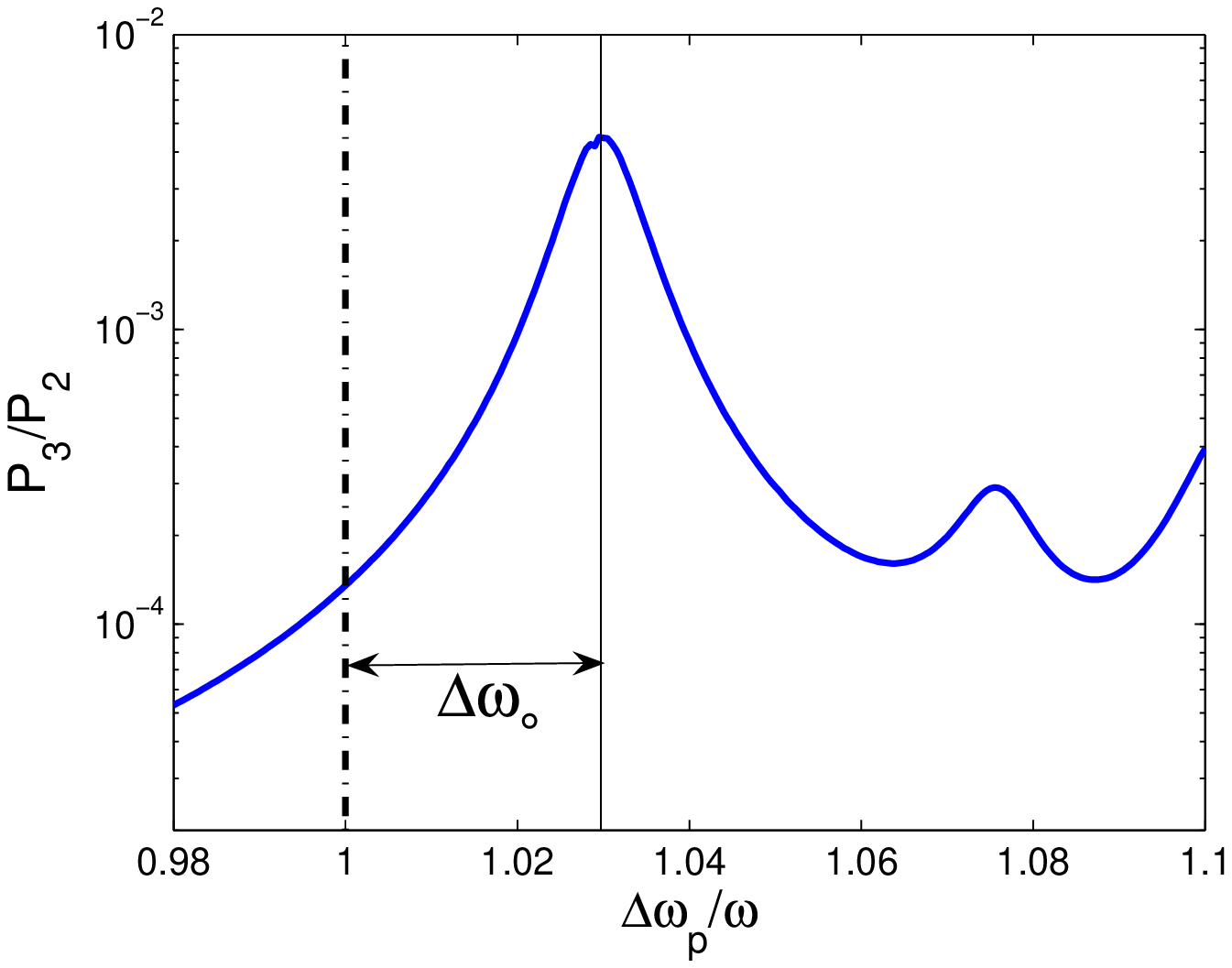}
\caption{Steady-state relative probability $P_3/P_2$ of three-particle excitation in the presence of a single delta potential localized at $r_\circ=0.4\ell$ under a monochromatic Laguerre-Gauss ${\rm LG}_0^3$ pump. System and pump parameters: $\Omega/\omega = 0.99$, $g_{nl}/\ell^2\omega = 4$, $\gamma/\omega = 0.01$, $\ell F/\gamma = 0.1$.\label{FigS5}}
\end{figure}

Some results for larger number of interacting particles are shown in Figs. \ref{FigS4} and \ref{FigS5}.

To obtain the braiding phase $\phi_{\rm Br}$ analytically for $N>2$ more efficiently, we used the relation (\ref{Berry_AngMom}). In Fig. \ref{FigS4}(a), we compare the estimation for $\phi_{\rm Br}$ obtained for $N = 3,4$ from the numerical ground state via Eq. (\ref{Berry_Energyshift}) with the analytical results from the quasi-hole wave functions. As compared to the $N=2$ case, it can be seen that the region of good agreement extends further in $r_\circ$, which could be expected as the cloud size increases with increasing $N$.

Panel (b) displays the analytical predictions for $\phi_{Br}$ as a function of $r_\circ$ for $N$ up to $6$; unfortunately, available computing resources prevent us from directly extending the calculation to higher $N$ values for which more sophisticated Monte Carlo techniques should be used. For increasing $N$, the position of maximum $\phi_{Br}$ shifts towards larger $r_\circ$ and its value gets close to $\pi$. This trend is quite different from that of the non-interacting system examined in the previous section, where the braiding phase is simply given by $N$ times the single-particle braiding phase.

For much larger number of particles, the characteristic incompressibility of the quantum Hall liquid is expected to result in a large plateau at $\phi_{Br}=\pi$. In this limit \cite{Wilczek_anyon,Yoshioka}, the Berry phase acquired through looping of a quasi-hole can simply be expressed as $2\pi\langle N\rangle$, $\langle N\rangle$ being the average number of particles enclosed by the loop. The effect of an additional quasi-hole inside the loop would be to displace a fraction $\nu$ of a particle from an otherwise spatially constant density, leading to an extra braiding phase of $2\pi\nu$. For the Laughlin and quasi-hole states considered in the present work, the fraction is $\nu=1/2$, which results in a braiding phase of $\pi$.

In Fig. \ref{FigS5} we simulate the optical excitation of an $N=3$ particle, one quasi-hole state using a Laguerre-Gauss ${\rm LG}_0^3$ pump with three units of angular momentum. At the position of the transmission peak at $\Delta\omega_p/\omega \approx 1.0297$, the overlap between the three-photon amplitude and the one quasi-hole wave function has the excellent value of $98.9\%$. The Berry phase calculated from the frequency shift $\Delta\omega_{\circ}$ of the peak via (\ref{Berry_Energyshift}) is $\approx0.163\pi$. This value is very close to the analytical prediction $\approx0.160\pi$ obtained from the analytical form of the one quasi-hole wave function, Eq. (3) of the main text, in the three-photon case. This suggests that our scheme to measure the many-body braiding phase can successfully extend to states with a larger number of particles.